\documentclass[preprint,amsmath,amssymb, aps, letterpaper]{revtex4} 
\usepackage{tabularx}
\usepackage{graphicx}
\usepackage{bm}
\usepackage{color}
\usepackage{float}

\newcommand{\be}{\begin{equation}}
\newcommand{\ee}{\end{equation}}
\newcommand{\bfig}{\begin{figure}}
\newcommand{\efig}{\end{figure}}
\newcommand{\incl}{\includegraphics}
\DeclareMathOperator{\sech}{sech}

\begin{document}

\title{Electronic Transport on the Shastry-Sutherland Lattice in Ising-type Rare Earth Tetraborides}

\author{Linda Ye}
\affiliation{Department of Physics, Massachusetts Institute of Technology, Cambridge, MA 02139}

\author{Takehito Suzuki}
\affiliation{Department of Physics, Massachusetts Institute of Technology, Cambridge, MA 02139}

\author{Joseph G. Checkelsky}
\affiliation{Department of Physics, Massachusetts Institute of Technology, Cambridge, MA 02139}

\date{\today}
\begin{abstract}
In the presence of a magnetic field frustrated spin systems may exhibit plateaus at fractional values of saturation magnetization.  Such plateau states are stabilized by classical and quantum mechanisms including order-by-disorder, triplon crystallization, and various competing order effects. In the case of electrically conducting systems, free electrons represent an incisive probe for the plateau states. Here we study the electrical transport of Ising-type rare earth tetraborides $R$B$_4$ ($R=$Er, Tm), a metallic Shastry-Sutherland lattice showing magnetization plateaus. We find that the longitudinal and transverse resistivities reflect scattering with both the static and dynamic plateau structure.  We model these results consistently with the expected strong uniaxial anisotropy in a quantitative level, providing a framework for the study of plateau states in metallic frustrated systems.    
\end{abstract}

\maketitle                  

\section{Introduction}
Geometrically frustrated lattices play host to a number of emergent quantum mechanical phases including quantum spin liquids \cite{QSL}, resonating valence bonds states \cite{RVB}, and complex magnetic orders \cite{complex}.  Such systems are typically electronic insulators constructed from low connectivity lattices that enforce competing magnetic interactions and enhanced quantum mechanical fluctuations \cite{RamirezReview}.  While in many cases introduction of charge carriers destabilizes such lattice-borne frustration, recently a variety of frustration-related effects have been discussed in this context in a class of materials termed frustrated metallic systems \cite{Review}.  Examples include kagome lattice model realizations of the fractional quantum Hall effect \cite{XG} and superconductors with exotic pairing symmetries \cite{BJ-SC, NT-SC}.  To what extent such phenomena can be realized in experiment is an open question.

\bfig[htb]           
\incl[width=0.8\columnwidth]{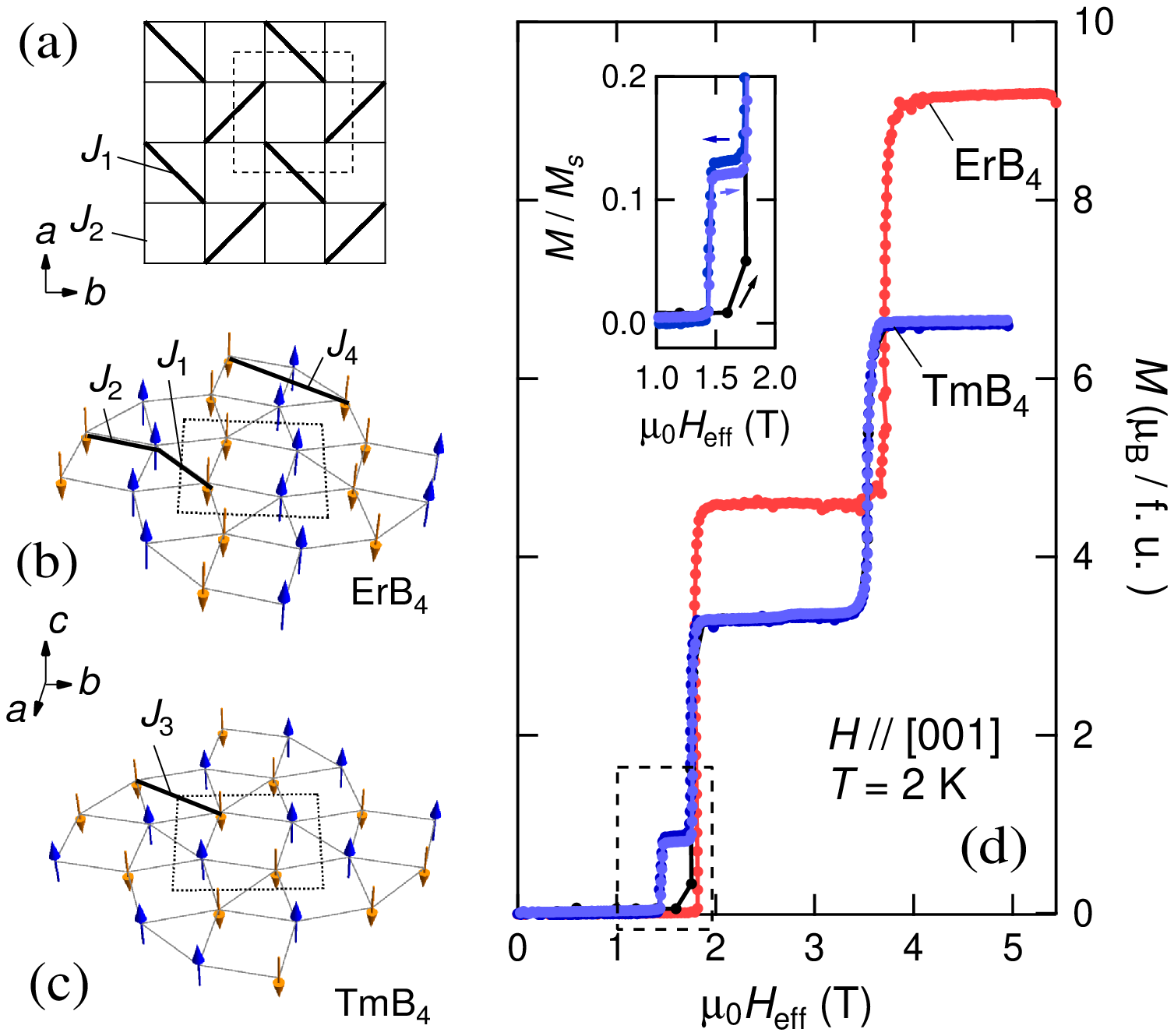}
\caption{\label{Fig1}
\textbf{Shastry-Sutherland Lattice and Magnetization plateaus in ErB$_4$ and TmB$_4$} (a) SSL model with diagonal bond $J_1$ and square bond $J_2$. (b), (c) Spin configuration for antiferromagnetic ground states in ErB$_{4}$ and TmB$_{4}$, respectively.  Exchange couplings $J_{1}$, $J_{2}$, $J_{3}$, $J_{4}$, and the unit cell (dashed line) are shown.  (d)  Magnetization as a function of field $\mu_0H_{\text{eff}}$ applied along the $c$-axis for ErB$_{4}$ and TmB$_{4}$.  The inset shows a magnified view near the $(1/q) M_s$ phase in TmB$_{4}$.}
\efig

A known materials system that has both lattice frustration and itinerant electronic behavior is the rare earth ($R$) tetraboride $R$B$_{4}$. The system is tetragonal (space group $P4/mbm$) with magnetic $R$ ions in the $ab$ plane forming a lattice topologically equivalent to the Shastry-Sutherland lattice (SSL) shown in Fig. \ref{Fig1}(a). While the 4$f$ electrons of the $R$ ions are localized in a frustrated configuration, the 4$d$ electrons from $R$ and 2$p$ from B act as itinerant carriers \cite{YB4calc}.  As with other SSL systems, the key parameters determining the frustration are the antiferromagnetic exchange $J_{1}$ and $J_{2}$ ($J_1,J_2>0$) on diagonal and square bonds on alternating tiles \cite{SS}.  Unlike the celebrated case of quantum spin-1/2 Cu$^{2+}$ ions in the insulating compound SrCu$_{2}$(BO$_{3}$)$_{2}$ which realizes the collective dimer singlet ground state predicted for the SSL \cite{SS,Cu1}, $R$B$_{4}$ has large classical $f$ moments with magnetic interactions mediated by itinerant electrons.  Despite this, just as SrCu$_{2}$(BO$_{3}$)$_{2}$ exhibits a series of fractional magnetization plateaus as a function of magnetic field $H$ with $M/M_{S} = 1/n$ ($n$ is an integer from 2 to 9, $M$ is the magnetization, and $M_{S}$ is the saturation $M$) \cite{Cu1,Cu2, Cu3, Cu4}, $R$B$_{4}$ also shows magnetization plateaus of unusual structure \cite{ErB4_MH,TmB4_neutron,TbB4_MH,ErTmHo}.  A particularly interesting limit is the trivalent $R = $ Er and Tm where a strong Ising single ion anisotropy exists such that the $f$-electron moments may be described as effective spin-1/2 moments locked perpendicular to SSL plane and the plateau transitions arise from complex spin flip processes \cite{RB4theory,superSolid,TmB4neutron}.

Herein we investigate how static and dynamic aspects of the magnetism in Ising-like $R$B$_{4}$ influence transport and the view it offers in to the energetics of the classical SSL magnetic phase diagram.  The SSL network for ErB$_{4}$ and TmB$_{4}$ along with their Ising-type antiferromagnetic (AFM) ground states are shown in Fig. \ref{Fig1}(b) and (c), respectively \cite{ErB4_neutron,TmB4_neutron}.  One view of the difference between the two systems is the connectivity of the spins: in ErB$_{4}$ the spins on the diagonal bonds are anti-parallel while in TmB$_{4}$ they are parallel.  This can be understood in terms of exchange interactions, as while both compounds have $J_1\approx J_2>0$ they differ in further neighbor interactions \cite{TmB4_neutron,superSolid}.  With $H \parallel c$, in ErB$_4$ the possible sites for field-dependent spin flips occur on 1D ferromagnetic chains connected by $J_2$ that are decoupled unless a fourth neighbor interaction $J_4$ is included.  For TmB$_{4}$ a third neighbor interaction $J_{3}$ complementary to $J_{1}$ allows instead for a 2D network of possible spin flips.  These differences can be connected to the corresponding plateau structures, which are shown in Fig. \ref{Fig1}(d).  Common to both systems are plateaus at $M_{S}/2$ while TmB$_{4}$ shows an additional plateau with higher denominator \cite{TmB4_neutron}.  As we discuss below, these differences in magnetism also have a significant impact on electronic transport.

\section{Methods}

Single crystals of ErB$_{4}$ and TmB$_{4}$ were grown using the floating zone method.  We reacted 99.99\% pure Er$_2$O$_3$ or 99.99\% pure Tm$_2$O$_3$ with 99\% pure B in Ar flow to form polycrystalline tetraborides \cite{ErB4_neutron}, from which single crystals were obtained after further zone refining.  Powder X-ray diffraction was done to confirm the materials are of a single phase and single crystal scattering was performed to orient crystals.  

Measurements of $M$ were performed using a commercial SQUID magnetometer.  The demagnetization factor $N$ calculated from sample dimensions \cite{Chikazumi} and the measured $M$ were used to obtain the effective field $H_{\text{eff}} = H - NM$ and magnetic induction $B = \mu_{0}(H_{\text{eff}} +M)$ for magnetization and transport measurements, respectively. Here $\mu_0$ is the vacuum permeability. The contributions from $R$ moments are significant with $\mu_0M_s=2.14$ T and 1.56 T for ErB$_4$ and TmB$_4$, respectively.  

Electrical measurements were performed using a standard low frequency (18.3 Hz) AC technique with a 2 mA excitation in a commercial cryostat. The dimensions of transport samples used here are  $0.71\times0.33\times0.02$ mm$^3$ (ErB$_4$) and $0.71\times0.28\times0.03$ mm$^3$ (TmB$_4$). $\rho_{xx}$ ($\rho_{yx}$) is obtained from symmetrization (anti-symmetrization) between time-reversed processes.

\section{Results and Discussion}

$R$B$_{4}$ are metals and the metallicity of ErB$_{4}$ and TmB$_{4}$ is similar.  Starting with ErB$_{4}$, as shown in Fig. \ref{Fig2}(a) the resistivity $\rho$ as a function of $T$ is metallic over the range $T = 2$ K to 300 K. There is a kink in $\rho(T)$ observed at low $T$ which corresponds to the AFM ordering temperature $T_{N}$ as observed in the temperature dependence of the magnetic susceptibility $\chi(T)$ shown in Fig. \ref{Fig2}(b).  The response is distinct from the shoulder-like features observed for typical antiferromagnetic metals such as Cr and Dy \cite{Meaden_Review}, where the antiferromagnetic ordering opens superzone gaps on the Fermi surface. Here this indicates an absence of Brillouin zone folding consistent with the AFM magnetic unit cell being identical to the crystallographic unit cell. The field-temperature phase diagram is shown in Fig. \ref{Fig2}(c); with increasing $\mu_{0}H_{\text{eff}}$ ErB$_{4}$ realizes a plateau state with $M_{S}/2$ and eventually enters a field-induced paramagnetic (FIP) phase (see also Fig. \ref{Fig1}(d)).  As shown in Fig. \ref{Fig2}(e), below $T_{N}$ a series of magnetoresistance features appear at the phase boundaries in Fig. \ref{Fig2}(c).  In particular, prominent peaks are observed at the magnetic transitions at moderate $T$ but are suppressed at the lowest $T = 2$ K. 

\bfig[tb]           
\incl[width=0.8\columnwidth]{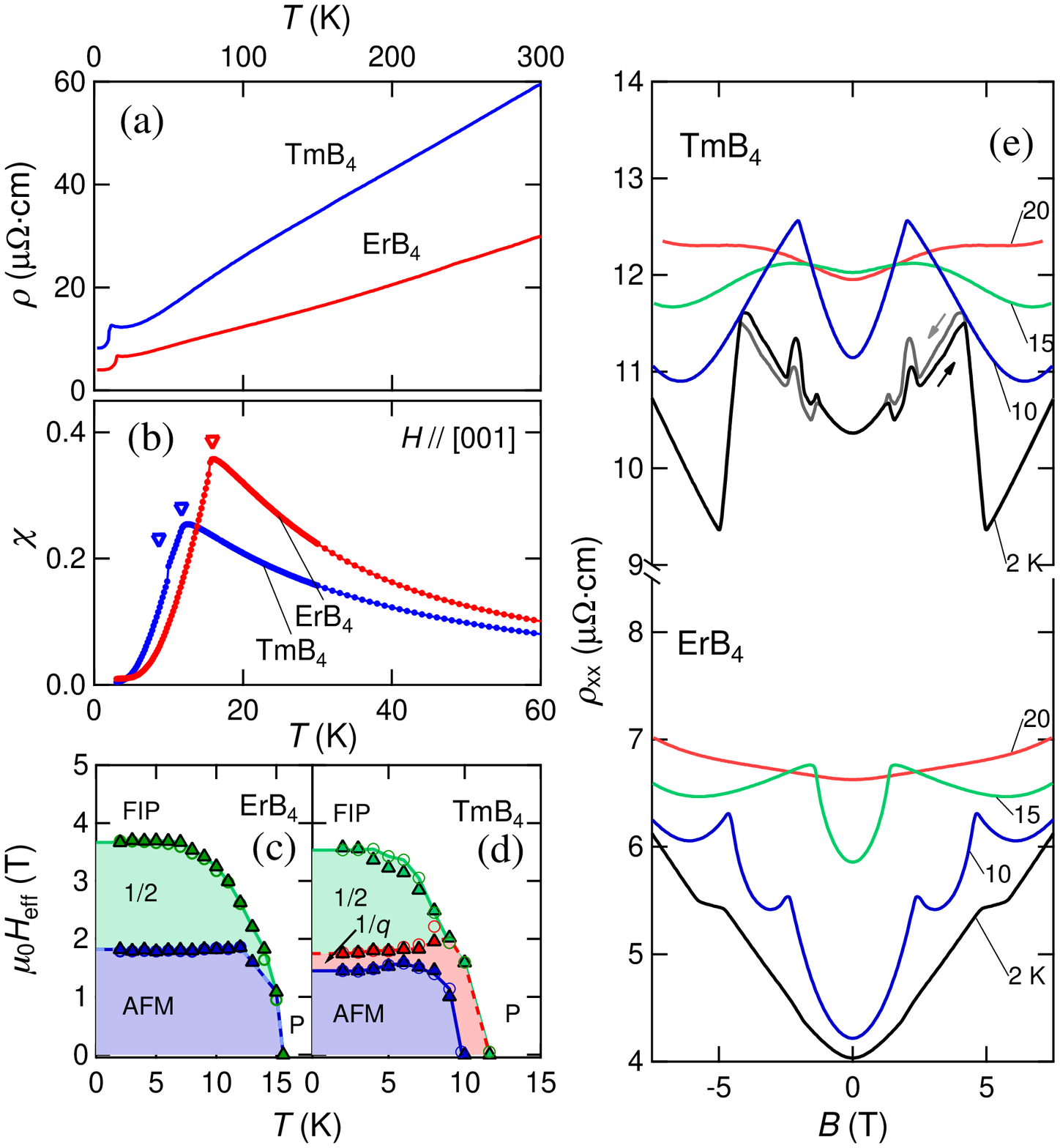}
\caption{\label{Fig2}
\textbf{Magnetic Phase Diagram of ErB$_4$ and TmB$_4$} (a) Resistivity $\rho$ as a function of temperature $T$ for the SSL plane of ErB$_{4}$ and TmB$_{4}$ single crystals. (b) Volume magnetic susceptibility $\chi$ measured along the $c$-axis for ErB$_{4}$ and TmB$_{4}$.  Triangles denote transition temperatures. (c), (d) Phase diagram in $H-T$ plane for ErB$_4$ and (trained) TmB$_4$, respectively. The boundaries determined from transport are shown with triangles and those from magnetization with circles.  (e) Magnetic field dependence of longitudinal resistivity $\rho_{xx}(B)$ at selected $T$ for TmB$_4$ and ErB$_4$.}
\efig

The overall behavior of TmB$_{4}$ is similar to that of ErB$_{4}$, but with an additional magnetic transition observed in $\rho(T)$ and $\chi(T)$ (Fig. \ref{Fig2}(a) and \ref{Fig2}(b), respectively) resulting in the phase diagram shown in Fig. \ref{Fig2}(d).  We denote the additional intermediate phase as $1/q$ as the value of $M$ in this region has been reported be history dependent ($q$ may take values of 7,9, or 11 \cite{TmB4_neutron}) and may not be precisely quantized \cite{TmB4_M_theory}.  Interestingly, this higher degree of complexity is also reflected qualitatively in $\rho_{xx}(B)$.  As shown in Fig. \ref{Fig2}(e), a low temperature hysteresis is observed in addition to sharp features corresponding to the magnetic transition.  

\subsection{Magnetoresistance in ErB$_4$} 
Detailed study of $\rho_{xx}(B)$ below $T_{N}$ reveals connections to the magnetic phases and transitions in the Ising SSL system.  We first focus on ErB$_{4}$ with $\rho_{xx}(B)$ shown in Fig. \ref{Fig3}(a).  The response can be understood as the sum of a conventional orbital magnetoresistance with additional scattering due to magnetic disorder and spin excitations as the plateau state evolves in field.  To isolate the magnetic contribution, we calculate $\Delta \rho_{xx}(B,T) \equiv \rho_{xx}(B,T) - \rho^{N}_{xx}(B,2$ K$)$, where we approximate the non-magnetic contribution $\rho^{N}_{xx}(B,2$ K$)$ (dashed line in Fig. \ref{Fig3}(a)) as a second order polynomial fit to the AFM and FIP phases where $M$ is constant at low $T$.  As shown in Fig. \ref{Fig3}(b), $\Delta \rho_{xx}$ exhibits a series of peaks at elevated $T$ and a residual enhancement at intermediate $B$.  

\bfig[tb]           
\incl[width=0.65\columnwidth]{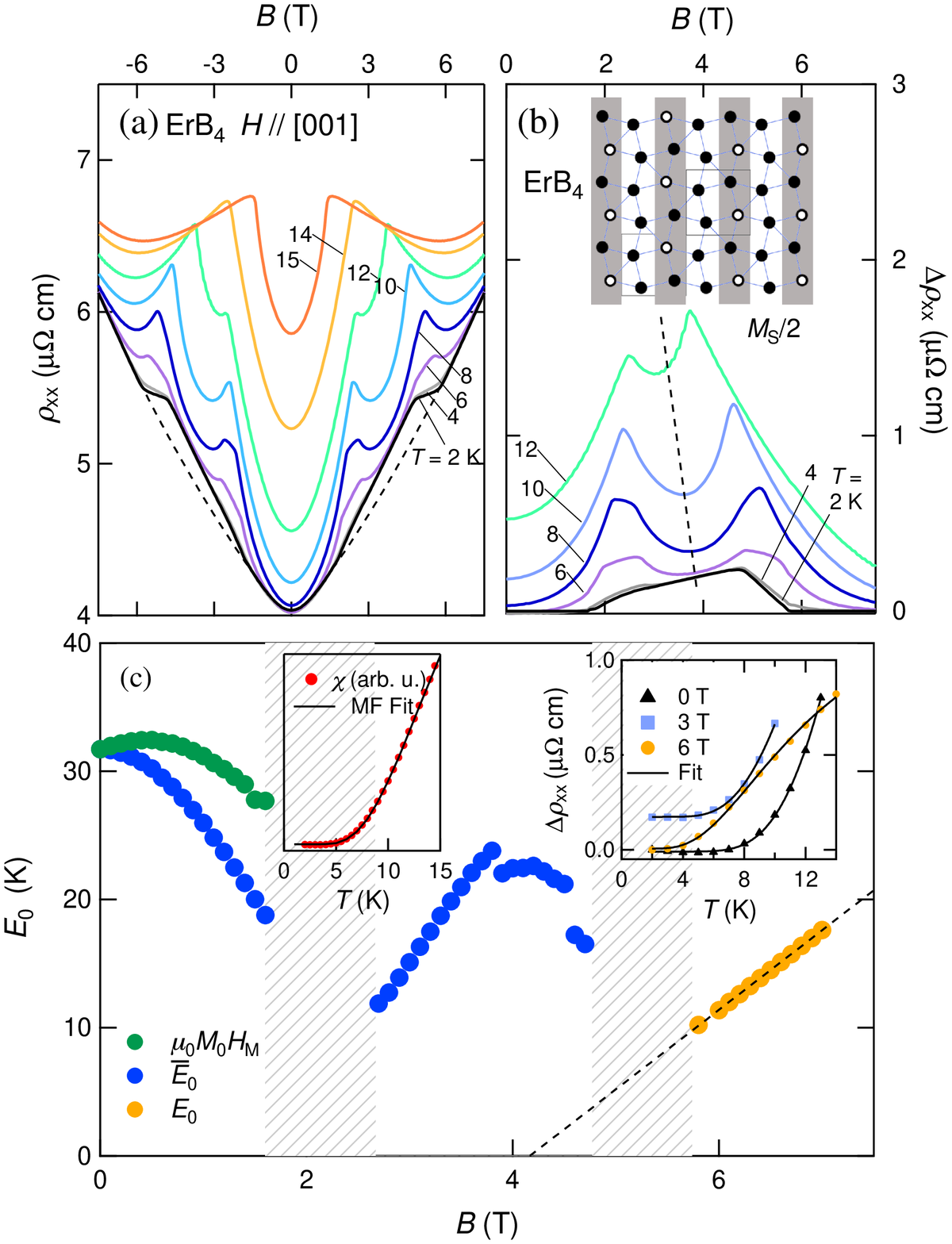}
\caption{\label{Fig3} \textbf{Magnetic scattering in ErB$_4$} (a) Detailed magnetic field dependence of longitudinal resistivity $\rho_{xx}(B)$ of ErB$_4$ at $T<T_N$. (b)  Magnetic contribution to resistivity $\Delta\rho_{xx}$ (see text). The inset shows a possible configuration of the half plateau state with the AFM stripes marked in gray. Here the black (white) circles represent spins parallel (anti-parallel) to $H$. The dashed square frames enclose two types of unit cells for the $M_s/2$ phase. (c) Magnetic energy $E_0$ as a function of $B$. The hatched area represents regions with phase coexistence and the dashed line is a linear fit for the FIP phase. The left inset shows a mean field fitting to the magnetic susceptibility of ErB$_4$ and the right inset shows fitting of $\Delta \rho_{xx}(T)$ at selected $B$.}  
\efig

Unlike $B$-induced changes in resistivity for $\rho^{N}_{xx}$ due to the Lorentz force, those in $\Delta \rho_{xx}$ arise from interaction of carriers with the magnetic state and therefore reflect a change in carrier relaxation time $\tau$. The coexisting $f$ moments and conduction electrons interact via a contact exchange interaction $\mathcal{H}_{cf} = J_{\text{c-f}}\bm{s}\cdot\bm{S}$, where $\bm{s}$ is the conduction electron spin and $\bm{S}$ is the total spin of localized magnetic moments \cite{Elliot,Meaden_Review}.  It has been proposed that the $M_{S}/2$ state is comprised of alternating AFM and ferromagnetic stripes (see the inset of Fig. \ref{Fig3}(b)) where a large degeneracy of ordering of the AFM stripes exists \cite{superSolid}. Such an additional degree of freedom can be expected to increase irregularities in the spin structure and therefore also in the periodic potential seen by the charge carriers causing increased scattering.  This is consistent with the step-like rise seen in both the raw $\rho_{xx}(B)$ trace and the $\Delta \rho_{xx}(B)$ peak in the $M_{S}/2$ phase.

The pattern at elevated $T$ in Fig. \ref{Fig3}(b) suggests thermally enhanced magnetic scattering. For antiferromagnets in the strong Ising-limit (where exchange energy is less than anisotropy energy), the lowest magnetic excitations are spin flips as classical spin waves cost considerable anisotropy energies. In this context, the $T$-excitation of the spin flips causes an increase in the spin-disorder resistivity (see Appendix A) in the following form \cite{CEF_Resistivity,RNi4B_Resistivity}:
\begin{equation}
\rho_m(T)\sim \sech^2(E_0/k_BT)
\label{AFMFit}
\end{equation}
where $E_0$ represents the magnetic energy at each site and $k_{B}$ is the Boltzmann constant. At $B=0$, $E_0$ equals $\mu_0M_0H_M$ with $M_0$ the rare earth magnetic moment and $H_M$ the effective molecular field on each site, and we get $E_0=32$ K from fitting $\Delta\rho_{xx}(T)$ with Eq.(\ref{AFMFit}). This is comparable with $E_0=23$ K obtained from the mean field fitting to the magnetic susceptibility of the Ising moments in ErB$_4$ \cite{VanVleck} (fit shown in Fig. \ref{Fig3}(c) left inset):
\begin{equation}
\chi(T)=\dfrac{1-m^2(T)}{T+E_0(1-m^2(T))}+\chi_0
\end{equation}
where $m(T)$ stands for the solution of sublattice magnetization at each $T$ to $m(T)=\tanh[E_0m(T)/T]$. $\chi_0$ represents the residual susceptibility which is rarely $T$-dependent.

Eq. (\ref{AFMFit}) may be further modified to describe the effects of finite fields taking $E_0=\mu_0M_0|H_M\pm H_{\text{eff}}|$ and the sign depends on whether the magnetic moments align or anti-align with the applied magnetic field. The green circles in Fig. \ref{Fig3}(c) show the fit results of $\mu_0M_0H_M$ taking half of all spins are parallel and half anti-parallel to $H_{\text{eff}}$, where $\mu_0M_0H_M$ depends weakly on $B$ within $30\pm5$ K. Alternatively, we show the average $\overline{E}_0$ obtained by from assuming a single uniform $E_0$ using blue circles, and the evolution of $\overline{E}_0$ with $B$ is shown in Fig. \ref{Fig3}(c) with representative fits to Eq. (\ref{AFMFit}) shown inset.  As $B$ is increased and the magnetic state is destabilized we see a drop in $\overline{E}_0$ from the zero field value 32 K.  At the magnetic transitions (regions corresponding to transitions in $M(H_{\text{eff}})$ shown as hatches areas in Fig. \ref{Fig3}(c)) a mixed magnetic phase is likely to exist not captured by the present model \cite{FeCl2}.  On entering the $M_{S}/2$ phase we see a rise in $\overline{E}_0$ to approximately 25 K where the state is most stable before it decreases again as the system approaches the transition to the FIP.

In the FIP phase, all the magnetic moments are uniformly aligned with $B$ and $E_0=\mu_0M_0(H_{\text{eff}}-H_{\text{M}})$, with the Zeeman energy gain associated with the applied field overwhelming the antiferromagnetic interactions. Here we expect a linear $B$-dependence of $E_0$ as is observed for fit results in FIP phase (orange circles in Fig. \ref{Fig3}(c)). The slope yields $M_0=9.24$ $\mu_B$, quantitatively consistent with the magnetic moment of Er$^{3+}$ ($M_s=9.6$ $\mu_B$/Er). The positive intercept on $B$ implies that the underlying interaction of the system is antiferromagnetic, and the FIP phase is destabilized at magnetic fields below 4 T.\\

\subsection{Hall Resistivity of ErB$_4$} 
We next examine the transverse resistivity $\rho_{yx}$.  As shown in Fig. \ref{Fig4}(a), there is an overall electron-like response with weak kinks appearing as a function of $B$.  The magnetic phase boundaries from the phase diagram in Fig. \ref{Fig2}(c) are shown as dashed lines and closely track the features in $\rho_{yx}$.  These features can be understood by the magnetic modifications to $\tau$ introduced above for $\rho_{xx}$.  We employ a modified two-band model incorporating a field-dependent relaxation time $\tau(B)$ for the longitudinal conductivity $\sigma_{xx}$ 
\begin{equation}
\sigma_{xx}=\sum_i\sigma_{xx}^i=\sum_i\dfrac{n_ie\mu_i(\tau(B)/\tau_0)}{1+(\mu_iB)^2(\tau(B)/\tau_0)^2}
\label{sigmaXX}
\end{equation}
where $\sigma_{xx}^i, n_i, \mu_i$ are the conductivity, carrier density and mobility of each band, and $\tau_0$ is the zero field relaxation time at a given $T$. The total transverse conductivity $\sigma_{xy}$ is written as
\begin{equation}
\sigma_{xy}=\sum_i\sigma_{xx}^i\cdot(\mu_iB)\cdot(\tau(B)/\tau_0)\label{sigmaXY}
\end{equation}
 The ratio $\tau(B)/\tau_0$ as shown in Fig. \ref{Fig4}(b) is obtained from $\Delta\rho_{xx}$, viz. $\tau(B)/\tau(0) = \rho_{xx}(0,T)/[\Delta\rho_{xx}(B,T)+\rho_{xx}(B, 2$ K$)]$.  

\bfig[tb]            
\incl[width=0.65\columnwidth]{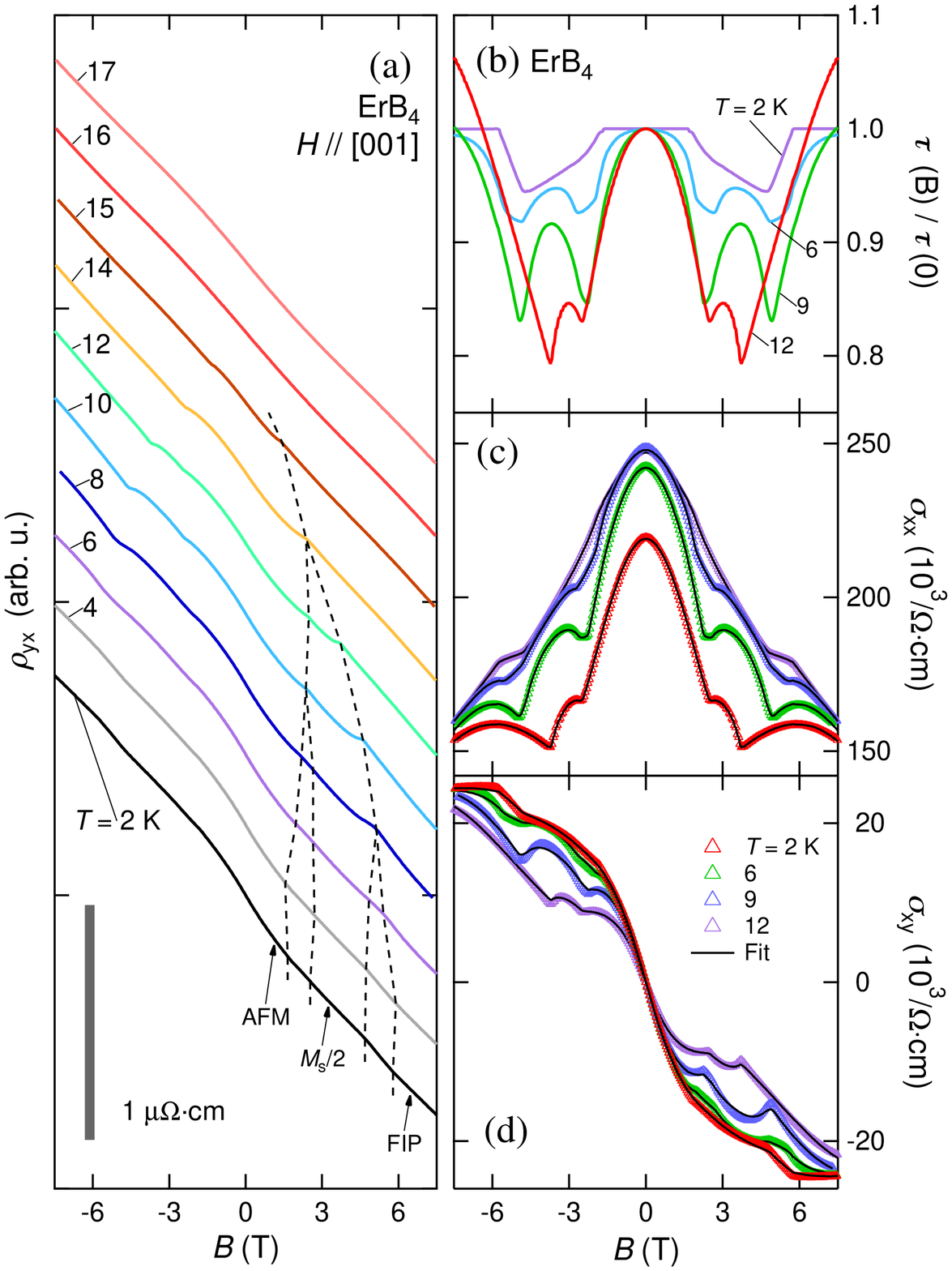}
\caption{\label{Fig4} \textbf{Hall effect in ErB$_4$} (a) Field dependence of transverse resistivity $\rho_{yx}(B)$ for ErB$_4$. The dashed lines represent the singularities observed in $\rho_{xx}$.  (b) Relative relaxation time $\tau(B)/\tau_0$ as a function of $B$. (c) Longitudinal conductivity $\sigma_{xx}$ and (d) transverse conductivity $\sigma_{xy}$ fit with modified two band model using $\tau(B)/\tau_0$ (Eqs. (\ref{sigmaXX}) and (\ref{sigmaXY})). The legend is the same for both panels.}
\efig

As shown in Figs. \ref{Fig4}(c) and \ref{Fig4}(d), Eqs. (\ref{sigmaXX}) and (\ref{sigmaXY}) provide satisfactory fits for $\sigma_{xx}$ and $\sigma_{xy}$, respectively.  The best fits for $\sigma_{xx}$ and $\sigma_{xy}$ at $T=2$ K are shown in Table \ref{Table1} (also for a second sample B). The set of parameters are similar for both fits, though there is a factor of 4-5 difference in carrier densities that optimize the longitudinal and transverse fits.  We hypothesize that the lack of convergence is related to the Fermi surface being composed of more than two bands \cite{YB4calc}.  However, higher ordering fitting is not a satisfactory proof of this given the large number of parameters it introduces.  

More generally, we suggest this demonstrates that the features in $\rho_{yx}$ may be captured by a field-induced scattering rate without showing clear signatures of anomalous Hall effect conventionally observed ferromagnets as a Hall effect proportional to $M$ \cite{NagaosaRMP}. We point out that the magnitude of anomalous Hall conductivity $\sigma_{xy}^A$ expected for the current system from the scaling relation between $\sigma_{xy}^A$ and $\sigma_{xx}$ is of the order $10^3$ /$\Omega\cdot$cm \cite{Universal}, which is difficult to unambiguously decompose from the background Hall conductivities  that shown prominent features upon magnetic phase transitions(see black fit curves in Fig. \ref{Fig4}(d)). We suggest that systems with reduced background $\sigma_{xy}$ from the normal Hall conductivity $\sigma_{xy}^N$ may provide a clearer view of the extrinsic/intrinsic anomalous Hall contributions in magnetization plateau systems. As $\sigma_{xy}^N\sim\tau$, this may be achieved by doping the boron sites in RB$_4$ with non-magnetic elements to suppress $\tau$ while minimizing the influence on the magnetic subsystem. Low carrier compounds are also favorable as they possess a smaller $\sigma_{xy}^N$ background though care must be taken as small carrier systems may exist at a different physical regime on the universal scaling \cite{Universal}.

\subsection{Transport in TmB$_4$}
Turning to the detailed magnetotransport of TmB$_{4}$, the low $T$  behavior of $\rho_{xx}$ and $\rho_{yx}$ are shown in Fig. \ref{Fig5}(a) and (b), respectively.  Unlike the case of ErB$_{4}$, we observe hysteresis in both transport channels (also recently reported in another study \cite{TmB4_canfield_transport}). Here hysteresis refers to the difference between time-reversed full field sweeps.  As shown in Fig. \ref{Fig1}(d) hysteresis is observed in $M(H)$ in the vicinity of the $(1/q)M_{s}$ phase; in transport hysteresis appears across a $B$ range corresponding to approximately both the $(1/q)M_{s}$ and $M_{s}$/2 phase.  Additionally, for $\rho_{xx}$ was observe a difference between the zero-field cooled (virgin) state and the trained state (that seen after once reaching the FIP phase).  

\bfig[tb]            
\incl[width=0.65\columnwidth]{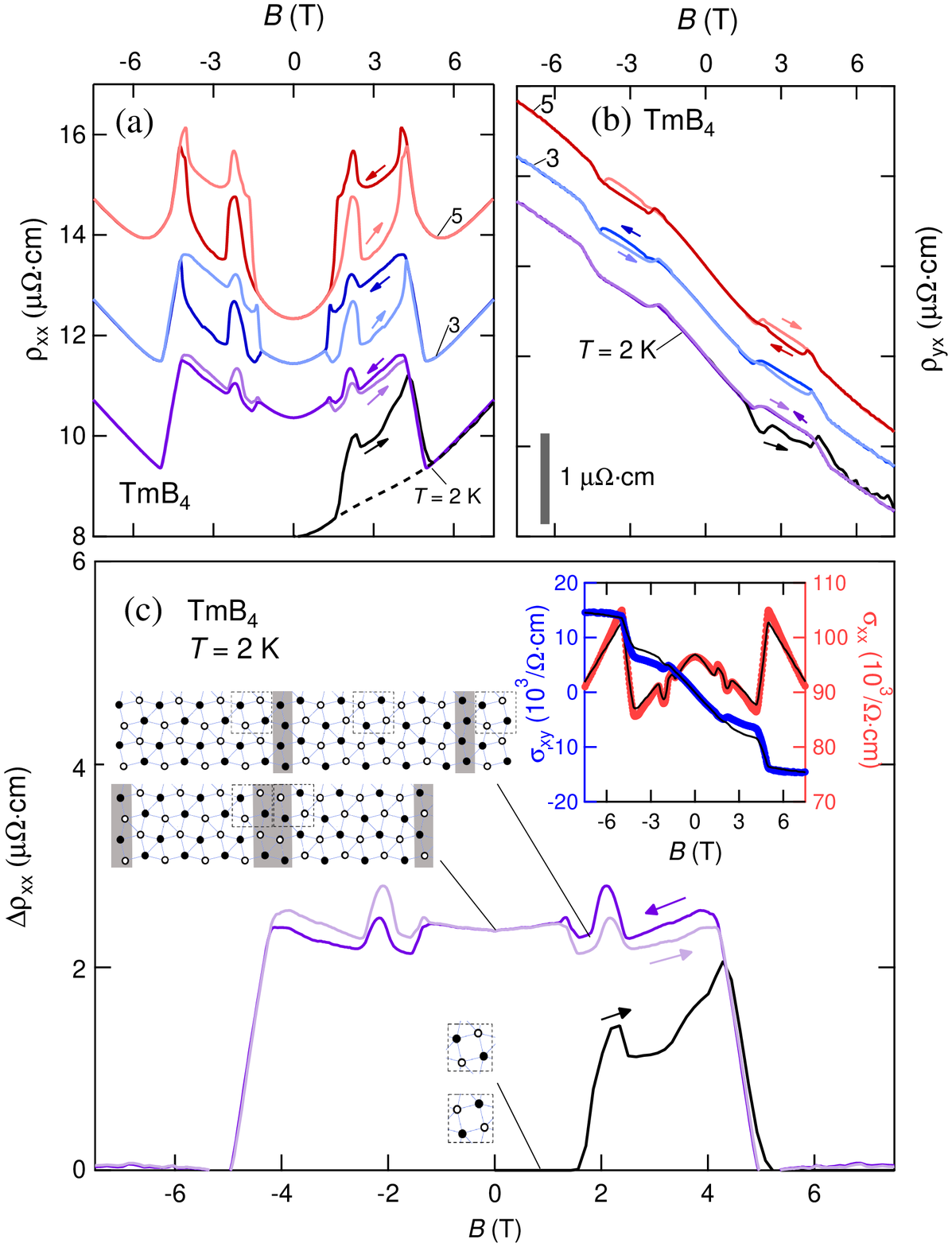}
\caption{\label{Fig5}
\textbf{Magnetotransport in TmB$_4$} (a) Longitudinal resistivity $\rho_{xx}$ for TmB$_{4}$ at low $T$.  The $T=$3 K and 5 K curves are offset for clarity.  (b) Transverse resistivity of TmB$_{4}$ at low $T$. (c) Magnetic contribution to resistivity $\Delta\rho_{xx}$ and possible magnetic configurations \cite{TmB4_neutron,TmB4_M_theory}. The inset shows $\sigma_{xx}$ and $\sigma_{xy}$ fit with Eqs. (\ref{sigmaXX}) and (\ref{sigmaXY}).  For clarity only scans from negative to positive $B$ are shown.}
\efig

To probe the origin of these effects, we construct $\Delta \rho_{xx}$ in a manner analogous to that for ErB$_{4}$.  In this case the normal component  $\rho_{xx}^N$ that connects the AFM and FIP states appears to belong to the virgin state, as shown with the dashed line in Fig. \ref{Fig5}(a).  Subtraction of this component yields $\Delta \rho_{xx}$ as shown in Fig. \ref{Fig5}(c).  The presence of additional scattering is evident in the trained phase.  We note that this is contrary to the case of conventional domain wall scattering in ferromagnets in which the virgin state typically has a higher resistivity \cite{Chikazumi}.  

Considerations of the detailed real space magnetic textures resulting from the 2D spin flip network in this system offer insight into this unusual behavior and more broadly the appearance of the $(1/q)M_s$ phase \cite{TmB4_M_theory}. The spin configuration for the zero-field cooled AFM state is known to have a magnetic unit cell identical to that of the crystallographic unit cell, as shown in Fig. \ref{Fig5}(c) \cite{TmB4_neutron}.  Starting from this simple AFM phase, with increasing $B$ the $M_{s}/2$ phase and then the FIP phase are stabilized.  Subsequent decreasing of $B$ to zero realizes a cascade of phases with $M=M_{s}/2$, $(1/q)M_{s}$, and 0.  However, these latter states are known to have larger real space magnetic structures, which are evidently nearly degenerate in energy and accessible along this thermodynamic path \cite{TmB4_neutron,TmB4_M_theory}.  One example of the expected long-period structure at $M=0$ is shown in Fig. \ref{Fig5}(c) with AFM domains in an anti-phase periodic structure.  It has been suggested that the alignment/shift of those AFM domains every 4/5 unit cells leads to the $(1/q)M_s$ phase in TmB$_4$ \cite{TmB4_M_theory}.  This characteristic of training and complexity is a hallmark of strong magnetic frustration in TmB$_{4}$; the resulting increase in $\rho_{xx}$ can then be viewed as due to domain wall scattering or the opening of superzone gaps in the Fermi surface if such structures are macroscopically ordered.  In contrast, time-reversal antisymmetric quantities $M$ and $\rho_{yx}$ do not show training.  

Similar to the case of ErB$_4$, the patterns observed in both $\rho_{xx}$ and $\rho_{yx}$ for TmB$_4$ can largely be explained by the magnetic structure-sensitive changes in $\tau$ and spin disorder in the plateau phases.  The fitting of $\sigma_{xx}$ and $\sigma_{xy}$ using Eqs. (\ref{sigmaXX}) and (\ref{sigmaXY}) is shown in the inset of Fig. \ref{Fig5}(c) (parameters are listed in Table \ref{Table2}).  Fitting of the transport reproduces the experimental curves apart from in the $M_s/2$ phase.  As deviations in the Hall response in magnetic systems are often due to the anomalous Hall effect, we suggest this may be due to a skew scattering contribution from the ferromagnetically aligned domain walls \cite{TmB4_M_theory,TmB4_neutron}.  In terms of modeling as employed in ErB$_{4}$, analysis of $\rho_{xx}(T)$ in the FIP phase yields a magnetic moment 6.84 $\mu_B$ ($M_s=$ 6.66 $\mu_B$/Tm), molecular field 1.74 T, and corresponding exchange energy -0.69 meV.  Here again transport offers a quantitative measure of the underling energy scales for the SSL. 
\section{Conclusion}
The present study demonstrates that transport is a sensitive probe of magnetic disorder and excitations in model metallic frustrated systems.  In particular, the magnetotransport processes are found to be sensitive to static and dynamic magnetic disorder across plateau transitions and allow for quantitative characterization of the underlying magnetic order and its excitations. These results are consistent with the strong Ising anisotropy expected for $R$ = Er and Tm. The results provide a framework to study the more complex $R$B$_{4}$ magnetization plateau series such as TbB$_4$ \cite{TbB4_MH} and HoB$_4$ \cite{ErTmHo} with non-Ising type anisotropies.  More broadly, our study offers a new approach to a central question in frustrated magnetic systems, \emph{i.e.} the nature of their elementary excitations.  Yb$_2$Pt$_2$Pb is a metal recently identified as an anomalous quasi 1D quantum magnet in which electronic transport may be a probe of spinon dynamics \cite{Yb2Pt2Pb}.  Further application to systems with novel excitations such as monopoles in spin ice \cite{SpinIce_Monopole}, spinons in spin liquids \cite{SpinLiquid} and quasi 1D quantum magnets \cite{1DSpinon} could offer new insights in to these phenomena.  

\begin{acknowledgments}
We are grateful to T. Senthil, I. Sodemann and T. Kurumaji for fruitful discussions. This research is funded in part by the Gordon and Betty Moore Foundation EPiQS Initiative, Grant GBMF3848 to J.G.C.  A portion of this work was performed at the National High Magnetic Field Laboratory, which is supported by National Science Foundation Cooperative Agreement No. DMR-1157490, the State of Florida, and the U.S. Department of Energy.  L.Y. acknowledges support by the STC Center for Integrated Quantum Materials, NSF Grant No. DMR-1231319 and by the Tsinghua Education Foundation.
\end{acknowledgments}

\section*{APPENDIX}

\subsection{Resistivity Analysis in ErB$_4$} 

We attribute the $T$-evolution of $\rho_{xx}$ to the inelastic scattering of conduction electrons by the magnetic subsystem. Due to the strong Ising anisotropy, the local moments can be adequately viewed as individual two-level systems splitted by molecular exchange fields. The level splititng is given by $2E_0=2\mu_0H_1M_0$. 

The contribution to resistivity from inelastic scattering on localized quantum levels can be modeled as (following the description of crystal field scattering \cite{CEF_Resistivity}):
\begin{equation}\label{CEF}
\rho\sim\dfrac{1}{\tau}\sim |J_{\text{c-f}}|^2\sum_{i,i'}|\langle m_s',i'|\bm{s}\cdot\bm{S}|m_s,i\rangle|^2p_if_{ii'}
\end{equation}
where $i$ and $i'$ ($m_s$ and $m_s'$) denote the initial and final states of the mangeitc moments (conduction electron spin), respectively. We define the occupation probability of the $i$-th level as $p_i$ and the Fermi factor as $f_{ii'}$ where
\begin{equation}
p_i=\dfrac{e^{-E_i/k_BT}}{\sum_j e^{-E_j/k_BT}}, f_{ii'}=\dfrac{2}{1+e^{(E_{i'}-E_{i})/k_BT}}
\end{equation}
Here $E_i$ and $E_{i'}$ are the energy of the localized moments before and after the scattering event, respectively.

Using $\pm$ to denote the two local levels with energies $\pm E_0$ we get
\begin{equation}\label{probability}
p_{\pm}=\dfrac{e^{\mp E_0/k_BT}}{e^{E_0/k_BT}+e^{-E_0/k_BT}}, 
\end{equation}
and the Fermi factor raising (lowering) the energy of the magnetic system is:
\begin{equation}\label{probability1}
f_{\mp,\pm}=\dfrac{2}{1+e^{\pm2E_0/k_BT}}=\dfrac{2}{e^{\mp E_0/k_BT}(e^{E_0/k_BT}+e^{-E_0/k_BT})}
\end{equation}
Substituting Eq. (\ref{probability}) and Eq. (\ref{probability1}) into Eq. (\ref{CEF}) we obtain the $T$-dependence of $\rho$ being
\begin{equation}
\rho \sim \sech^2(E_0/k_BT)
\end{equation}
\newline

\subsection{Parameters for Two-band Fitting} 

For ErB$_4$ we fit eqn.(\ref{sigmaXX}) and (\ref{sigmaXY}) to $\sigma_{xx}$ and $\sigma_{xy}$ of two samples A and B respectively. Below in Table \ref{Table1} we show the fitting parameters for sample A at 2 K up to 9 T, and for sample B at 1.6 K up to 18 T. In each case there exist two electron bands with relatively high (low) density and low (high) mobility.

\begin{table}[h]
\renewcommand{\arraystretch}{1}
\caption{Fitting parameters for two-band model for ErB$_4$.  Sample A is measured between $\pm$ 9 T while sample B is measured between $\pm$ 18 T.}
\begin{tabular}{c c c c c}
\hline
\hline
 & $n_1$ (/cm$^3$) & $\mu_1$ (cm$^2$/V$\cdot$s) & $n_2$ (/cm$^3$) &  $\mu_2$ (cm$^2$/V$\cdot$s) \\
\hline
$\sigma_{xx}$ (A) & 1.74$\times10^{21}$ & 716.9 & 8.15$\times10^{19}$ & 3680.5 \\
$\sigma_{xy}$ (A) & 3.17$\times10^{20}$ & 1218 & 1.816$\times10^{19}$ & 4036 \\ 
$\sigma_{xx}$ (B) & 1.26$\times10^{21}$ & 484.3 & 2.04$\times10^{20}$ & 2064 \\
$\sigma_{xy}$ (B) & 2.2$\times10^{20}$ & 1186 & 1.83$\times10^{19}$ & 4441.7 \\
\hline 
\hline
\end{tabular}
\label{Table1}
\end{table}

For TmB$_{4}$, to avoid complications of the observed hysteresis we fit the negative to positive field scan with resulting parameters shown in Table \ref{Table2}. Similarly, two electron-like bands contributes to the conductivity.

\begin{table}[h]
\renewcommand{\arraystretch}{1}
\centering
\caption{Fitting parameters for two-band model for TmB$_4$ (scan with increasing $B$)}
\begin{tabular}{c c c c c}
\hline
\hline
 & $n_1$ (/cm$^3$) & $\mu_1$ (cm$^2$/V$\cdot$s) & $n_2$ (/cm$^3$) &  $\mu_2$ (cm$^2$/V$\cdot$s) \\
\hline
$\sigma_{xx}$ & 1.12$\times10^{21}$ & 624 & 3.95$\times10^{18}$ & 10467 \\
$\sigma_{xy}$ & 6.8$\times10^{19}$ & 1024 & 8.12$\times10^{19}$ & 1063 \\
\hline
\hline
\end{tabular}
\label{Table2}
\end{table}
\color{black}

\end{document}